\documentclass[12pt]{iopart}

\usepackage{iopams}  
\usepackage{epigraph}

\newcommand{\beq}{\begin{equation}}
\newcommand{\eeq}{\end{equation}}
\newcommand{\beqa}{\begin{eqnarray}}
\newcommand{\eeqa}{\end{eqnarray}}

\usepackage{graphicx}
\usepackage{cite}
\usepackage[table]{xcolor}
\usepackage{color}

\usepackage{calrsfs}
\DeclareMathAlphabet{\pazocal}{OMS}{zplm}{m}{n}

\begin{document}

\title{Energy consumption for ion transport in a segmented Paul trap}

\author{A. Tobalina}
\address{Department of Physical Chemistry, University of the Basque Country UPV/EHU, Apdo 644, Bilbao, Spain}
\ead{ander.tobalina@ehu.eus}
\author{J. Alonso}
\address{Institute for Quantum Electronics, ETH Z\"urich, Otto-Stern-Weg 1, 8093 Z\"urich, Switzerland}
\author{J. G. Muga}
\address{Department of Physical Chemistry, University of the Basque Country UPV/EHU, Apdo 644, Bilbao, Spain}
\vspace{10pt}

\begin{abstract}
There is recent interest in determining energy costs of shortcuts to adiabaticity (STA), 
but different definitions of  ``cost'' have been used.  
We demonstrate 
the importance of taking into account the Control System (CS) for a fair assessment of energy flows and consumptions. 
We model the energy consumption and power to transport an ion by a  STA protocol 
in a multisegmented Paul trap.       
The ion is driven by an externally controlled, moving harmonic oscillator. 
Even if no net ion-energy is gained at destination, setting the time-dependent control parameters is a macroscopic operation that 
costs energy and results in energy dissipation for the short time scales implied by the intrinsically fast STA processes.
The potential minimum is displaced 
by modulating the voltages on control (dc) electrodes. A secondary effect of the modulation, usually ignored 
as it does not affect the ion dynamics, is the time-dependent energy shift of the potential minimum.   
The non trivial part of the energy consumption 
is due to the electromotive forces to set the electrode voltages through the low-pass filters required to preserve the electronic noise from decohering the ion's motion.  
The results for  the macroscopic CS (the Paul trap) are compared to the microscopic power and energy of the ion alone. 
Similarities are found -and may be used quantitatively to minimize costs- 
only when the CS-dependent energy shift of the harmonic oscillator is included in the ion energy.    
\end{abstract}

%
%
%
%
%


$^{}$\vspace*{2cm}
\epigraph{To arrive at abstraction it is always necessary to begin with a concrete reality}{\textit{Pablo Picasso}\\ {\scriptsize{as quoted in 
``Conversations with Picasso", G. Brassa\"i,  University of Chicago Press, 1999}}}
\section{Introduction}
%
%
%
Several papers    
\cite{Campbell2017,Zheng2016, Abah2017a ,Funo2017,Chen2010b,DelCampo2014,Santos2016,Coulamy2016,Kosloff2017, Kieferova2014, Bravetti2017,Abah2017b}
have studied the ``energy cost'' or ``energy consumption''
of shortcuts to adiabaticity (STA)  \cite{Torrontegui2013, Chen2010a},
fast track routes to the results of slow adiabatic processes. 
Assessing the energy consumption of STA protocols is particularly relevant in quantum thermodynamics as they may appear to 
imply zero costs above the differential between initial and final energies, for example in expansion/compression  strokes of a quantum heat engine or refrigerator.      
Often the Primary System (PS), whose state is of interest  for the application at hand,
is microscopic while the Control System (CS) is macroscopic, so that 
the PS is described as governed by a semiclassical Hamiltonian with (classical) external time-dependent control parameters.  
Different STA are commonly formulated by specifying the protocol, i.e.,  the time dependences of the parameters that induce
fast state changes of the PS.

While  the cited  works ignore the energetic needs of  control elements and focus on 
the energy of the PS,  
or even on parts of the Hamiltonian of the PS,  in \cite{Torrontegui2017} a more general approach was suggested. There, the energy flow with the outer world is studied for an enlarged system that includes the PS  and the CS required to change the time-dependent parameters that drive the PS. The divide between the enlarged system PS+CS and the outer world should be drawn such that the energy flow through that boundary  can indeed be  translated into actual fuel or electric power consumption. 
For recent, related discussions of the need to include a CS along with the PS, see e.g. \cite{Horowitz2015}, where the energy required to manipulate a mesoscopic quantum system in the presence of noise is examined, or \cite{Clivaz2017}, where   fundamental limits of quantum refrigeration are discussed. 

Torrontegui et al. carried out their study  for a mechanical system that could be
thoroughly analyzed,  
the transport of a load (PS) suspended from a moving trolley (CS) in a mechanical crane
\cite{Torrontegui2017}. This model is in fact quite close mathematically  
to  experimental setups  that shuttle  ions or cold atoms by moving mirrors \cite{Zenesini2009} or lenses \cite{Couvert2008}. 
A number of conclusions that were
conjectured to be broadly applicable were drawn in \cite{Torrontegui2017}.
To test these conclusions and explore different models that may help to build general concepts 
and an embracing theory, it is worth investigating the energies involved in different transport   
experiments that do not rely on control elements subjected to mechanical displacements.   

\begin{figure}[h]
\begin{center}
\includegraphics[width=0.55\textwidth]{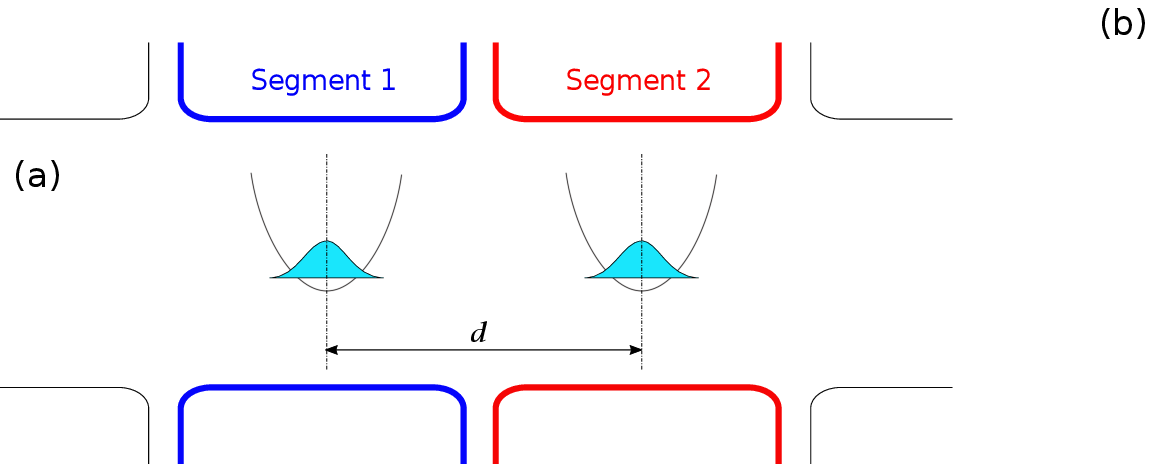}
\includegraphics[width=0.35\textwidth]{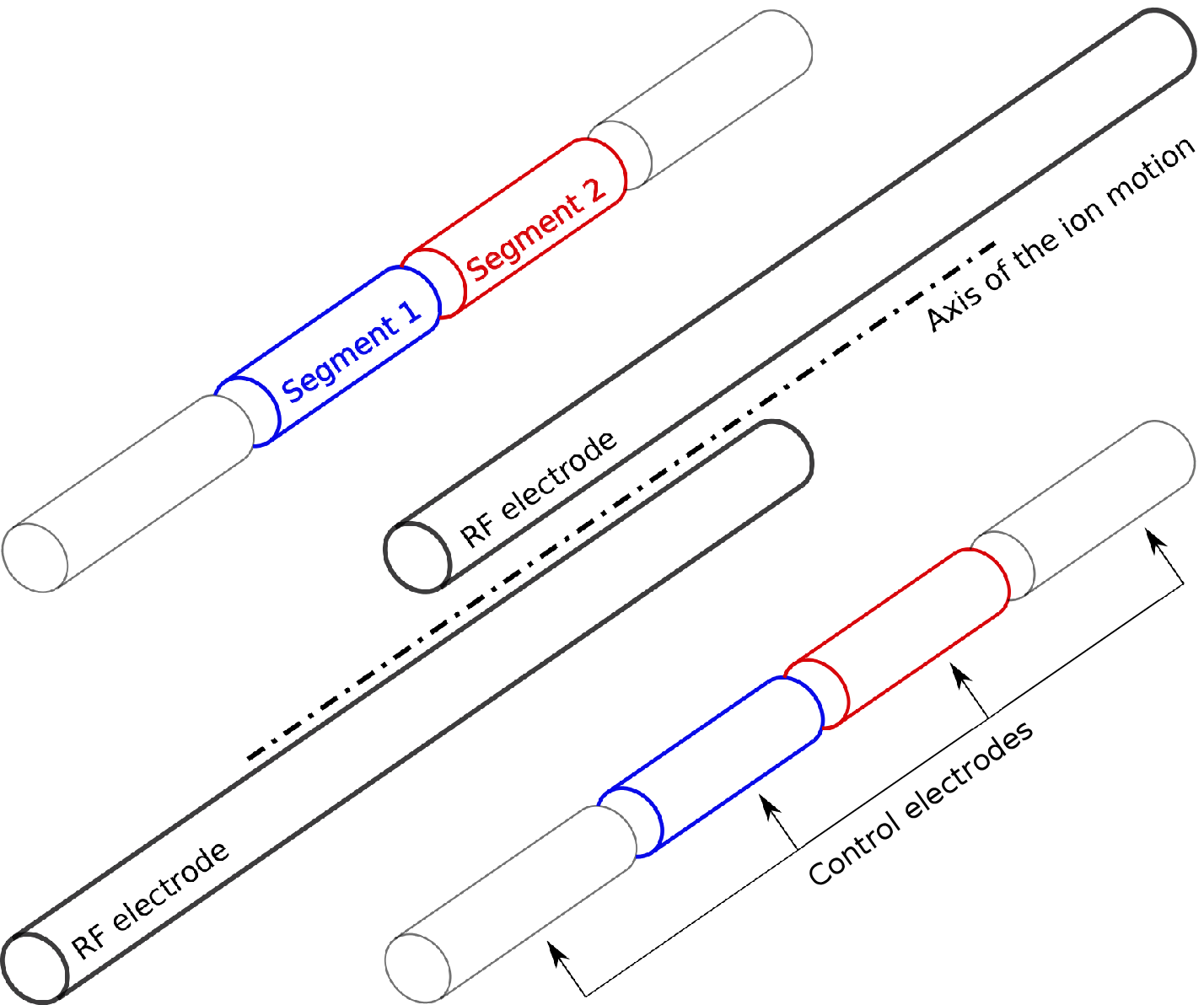}\vspace{1cm}
\includegraphics[width=0.45\textwidth]{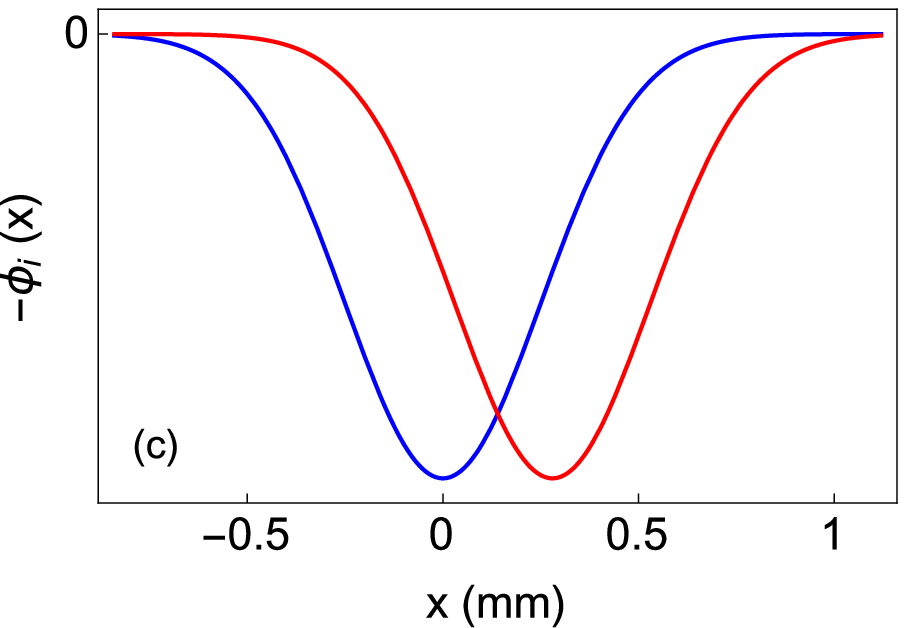}
\caption{(a) Schematic representation of the ion shuttling process and (b) layout of the electrodes of a segmented linear Paul trap. Segments of facing pair of electrodes in red and blue produce the potential that transports the ion while the rest remain grounded. RF electrodes provide axial confinement. (c) Electrostatic potentials modeled as $\phi_i=a \,e^{-(x-b_i)^2/c^2}$, where subindices $i=1,2$ correspond to the different segments. The parameters used for the Gaussian curves, chosen to fit the electrostatic potentials obtained in \cite{Furst2014}, are $a = 0.2$, $b_1=0$, $b_2=d$ and $c = 250$ $\mu$m, and will be used throughout the paper. The transport distance is  $d = 280$ $\mu$m. \label{gaussiancurves}}
\end{center}
\end{figure}

\begin{figure}[h]
\begin{center}
\includegraphics[width=0.7\textwidth]{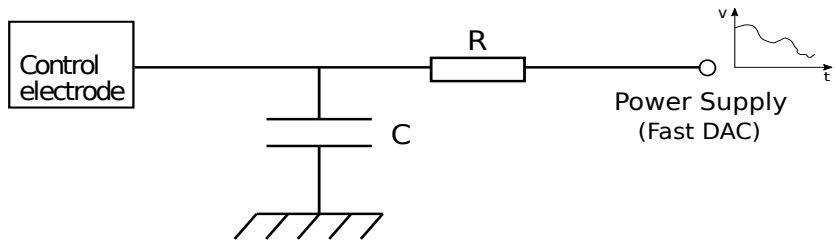}
\caption{Electronic scheme for setting the voltages at the control electrodes. It consists of a digital-to-analog converter (DAC) that allows for waveforms, and a low-pass electronic filter built with a resistor $R$ and a capacitor $C$.  \label{circuitdiagram}}
\end{center}
\end{figure}

Here we shall study the energy consumption to transport via STA a single ion (PS), in its ground motional state at initial and final times, in a 
linear  Paul trap made of parallel radiofrequency (rf) electrodes and segmented 
pairs of dc electrodes  \cite{Rowe2002,Leibfried2003a,Singer2010}, see figures \ref{gaussiancurves}(a) and \ref{gaussiancurves}(b). Strong radial confinement is assumed, 
which is primarily due to
ponderomotive forces caused by the rf field, whereas the potential along the longitudinal trap axis $x$ is controlled 
by the voltage biases applied to the control electrode pairs of each segment \cite{Wineland1998}. The potential minimum 
is displaced along the trap axis by applying waveforms that change the voltages of the control electrodes in time.     
Adiabatic \cite{Rowe2002} and faster-than-adiabatic shuttling experiments of this type have been performed  \cite{Bowler2012,Walther2012}. 
In our simplified model, and without loss of generality, we consider the transport of an ion between two nearby segments with centers at $x=0$ and $x=d$, as in \cite{Furst2014}. 
%
%
The voltage in each segment of facing pairs of dc electrodes is controlled by a programmable waveform generator and a low-pass electronic filter as shown in figure \ref{circuitdiagram}. The latter is used in trapped-ion experimental setups to limit the heating and decohering action of electronic noise on the ion motion. Filters are preferably placed close to the trap electrodes, inside the vacuum chamber housing the trap. In this way, it is possible to suppress significantly the amplitude of noise generated at the voltage supplies or picked up along the wires connecting these to the trap electrodes \cite{Alonso2013}. The filters are commonly built with a resistor $R$ and capacitor $C$ (first-order RC filters), although higher-order filters and active filters are also possible. 
In this work we will consider RC filters without incurring in loss of generality, since finite resistances and large capacitors are inherent to the control circuitry regardless of the filters used, whereas parasitic inductances produce negligible effects.
We assume a constant power 
 supply to generate the rf field, which makes this consumption trivial,
unlike that due to the voltage waveforms applied at the control dc-electrodes.   
In this model, the energy flow between the enlarged system implies a consumption of power due to energy dissipated by the resistances, 
and the energy required to charge and discharge the capacitors. 
In the mechanical analogy of  \cite{Torrontegui2017} different limits were identified depending on whether time intervals with negative power of the control consume energy,  save it, or become energetically neutral.  In the current model for the ion-transport process the capacitor  charge and discharge have to be actively driven, and thus both imply consumption.  This is analogous to the scenario in which both the accelerating and the braking phases of the control trolley use an engine to pull the trolley in different directions in the mechanical analogy.

The specific STA protocol we consider here to set the time-dependent location of the axial potential is based on 
the ``compensating force approach''. This technique compensates with a homogeneous, time-dependent force the inertial forces due to the motion of a reference trap trajectory, so the ion wave function remains at rest in the frame moving with the reference trap \cite{Torrontegui2011,Masuda2010,Ibanez2012}. It amounts to 
the trick that a waiter uses to carry  the tray quickly, tilting it 
to avoid spilling the drinks \cite{Sels2017}. In the harmonic approximation
for the trap, the compensation displaces the minimum.  
Within the set of STA-transport protocols based solely on choosing a certain path for the harmonic trap, the compensating force approach is generic in the sense that any reference trajectory is allowed, subjected to certain boundary conditions. The compensating force approach may be also regarded as invariant-based  inverse engineering of the transport protocol \cite{Torrontegui2011}, as explained in the next section.  
Other STA  transport protocols may be  based on counterdiabatic driving,
which changes the structure of the Hamiltonian adding a momentum dependent interaction \cite{Torrontegui2011}.  
The counterdiabatic (CD) driving method and the compensating force approach are unitarily connected -they can be found from each other
by a unitary transformation- \cite{Ibanez2012,An2016,Chen2017},
although the physical implementation involves different interactions and a different experimental  setting. Actual transport -in the fixed laboratory frame- has not yet been implemented with CD driving although An et al. \cite{An2016} simulated CD transport experimentally in an interaction picture with respect to the harmonic oscillation. They also performed the compensating force approach as ``unitarily equivalent transport'' in the interaction picture.   
The driving forces were induced optically rather than by varying voltages of control electrodes. 
Controllable momentum and spin dependent interactions for actual CD-driven transport in the lab frame may in principle be applied with synthetic spin-orbit coupling \cite{Chen2018} but the spin dependence would be a strong limitation for many applications, e.g. to transport arbitrary qubits.   
The corresponding energetic analysis  lays beyond the scope of this work.     


In Section \ref{methods} we review briefly the compensating force approach and find the voltages needed to implement the desired potentials.
This will also set the time-dependent term in the PS Hamiltonian. 
In Section \ref{results} we define and compare the different energies and powers involved. Power peaks that limit 
how short the process times may be, asymptotic dependences, and an optimization of the consumption are also discussed. 
The values of the parameters used in the computations have been taken from \cite{Furst2014}.
The paper ends with a summary and outlook for future work.   
\section{Methods\label{methods}}
%
%
%
%
%
\textit{Compensating force approach for a transport process}. 
Let us consider an ion of mass $m$ driven by a Hamiltonian of the form $H=\frac{p^2}{2m}+V(x,t)$, with 
\beq
\label{compensatingforcepotential}
V(x,t) = -F(t) x + \frac m 2 \omega^2 [x-\alpha(t)] ^2 + f(t),
\eeq
where 
\beq
\label{F(t)}
F(t)=m\ddot{\alpha}(t), 
\eeq
and dots represent derivatives with respect to time.  
$F(t)$ is a homogeneous force that compensates the inertial force generated by the acceleration of the reference harmonic
potential with angular frequency $\omega$ given by the second, quadratic term in (\ref{compensatingforcepotential})\cite{Torrontegui2011}. 
$\alpha(t)$ may be in principle an arbitrary reference trajectory from $\alpha(0)=0$ to $\alpha(t_f)=d$
in a given time $t_f$. 
Different trapping configurations, such us a non-rigid harmonic potential or a double well potential have been examined for more complex transport protocols, e.g. in \cite{Tobalina2017}, but these generalizations are not needed for our current purpose. 

$H$ supports an invariant of motion, 
$I=\left[p-m\dot{\alpha}(t)\right]^2/(2m) + \frac m 2 \omega^2 [x-\alpha(t)]^2$,
provided that the force $F(t)$ and $\alpha(t)$ satisfy (\ref{F(t)})
%
%
\cite{Torrontegui2013}.
Any wave function $\Psi(x,t)$ that evolves with $H$ may be expanded in terms of eigenvectors $\psi_n$ of $I$,
\beq
\Psi(x,t)=\sum_n c_n e^{i \theta_n} \psi_n (x,t), \hspace{1cm}    I(t)\psi_n(x,t)=\lambda_n \psi_n(x,t),
\eeq
where $c_n$ are constant coefficients, $\lambda_n$ the time-independent eigenvalues of the invariant, and $\theta_n$
are Lewis-Riesenfeld phases that can be calculated as \cite{Lewis1969} 
$\theta_n(t)=\frac{1}{\hbar}\int_0^{t} dt' \langle \psi_n(x,t')|i\hbar \frac{\partial}{\partial t}-H(t')|\psi_n(x,t')\rangle.$
The eigenstates of the invariant can be written as \cite{Dhara1984}
\beq
\label{inveigenstates}
\psi_n(x,t)=e^{\frac{im}{\hbar}\dot{\alpha}(t)x} \Phi_n[x-\alpha(t)],
\eeq
where $\Phi_n$ are the eigenfunctions of the harmonic oscillator centered at $\alpha(t)$, 
the ``transport function''. 
%
%
%

The purely time-dependent potential energy  term $f(t)$ in (\ref{compensatingforcepotential}) is
frequently ignored 
since it 
``only adds'' a global phase to the wave function \cite{Torrontegui2012}. 
Nevertheless, this term is physically meaningful.
In particular,  it will determine the actual energy of the ion relative to a fixed zero of energy  
and the corresponding power.

\begin{figure}[h]
\begin{center}
\includegraphics[width=0.5\textwidth]{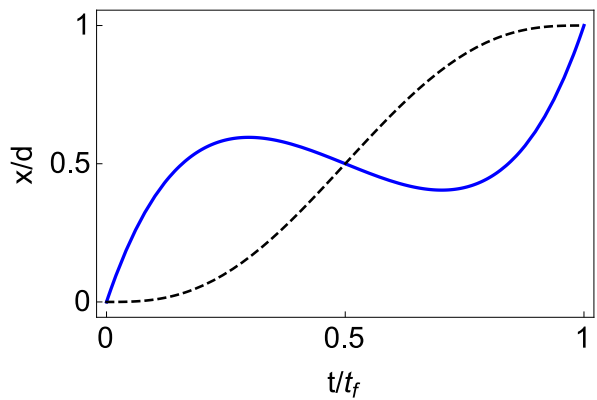}
\caption{Trajectory of the minimum of the potential (\ref{compensatingforcepotential})
(solid blue line), and $\alpha(t)$ in (\ref{alfa}) (dashed black line), which is the center of the dynamical states (\ref{inveigenstates}) and the minimum of the reference harmonic potential. The parameters used are $t_f=0.418$ $\mu$s, $d = 280$ $\mu$m and $\omega=2\pi\times1.3$ MHz. \label{trajectories}}
\end{center}
\end{figure}

The potential (\ref{compensatingforcepotential}) drives the ion from an initial to a non-excited displaced state
if we impose commutativity between the Hamiltonian and its invariant at boundary times and thus $H(t_b)$ and $I(t_b)$ share eigenstates ($t_b=0,t_f$). 
A simple choice for the transport function is $\alpha(t)=\sum_{j=0}^5 \alpha_j (t/t_f)^j$. While other functional forms are also possible, the polynomial function is known to yield smooth and technically feasible results \cite{Torrontegui2011}. The parameters $\alpha_j$ are fixed so that $\alpha(t)$ satisfies 
$\alpha(0)=0$, $\alpha(t_f)=d$, $\dot{\alpha}(t_b)=0$ for commutativity, and also $\ddot{\alpha}(t_b)=0$ to have a continuous force with $F=0$ for $t\le 0$ and $t\ge t_f$. These boundary conditions yield
\beq
\label{alfa}
\alpha(t) = d\left[10(t/t_f)^3-15(t/t_f)^4+6(t/t_f)^5\right].
\eeq
%
Unless stated otherwise, we shall use the transport function in (\ref{alfa}) in the examples and computations. Later in Section \ref{optim} we shall use a higher order polynomial 
with additional freedom to optimize consumptions. 
Note that $\alpha(t)$ represents the trajectory of the center of the dynamical states  (\ref{inveigenstates}),  which coincides with the minimum of the reference harmonic potential $\frac m 2 \omega^2 [x-\alpha(t)] ^2$, but not with the trajectory followed by the minimum of the total potential (\ref{compensatingforcepotential}), displaced due to the compensating force to $\alpha(t)+\ddot{\alpha}(t)/\omega^2$, 
as shown in figure \ref{trajectories}. 
%

%
%
%
%
%
%
%
%

\noindent\textit{Evolution of Segment Voltages}. 
We consider a simple setting to transport the ion between two (pairs of) electrodes centered at $x=0$ and $x=d$.  The time-dependent potential in (\ref{compensatingforcepotential}) that shuttles the ion is in practice generated as
a local approximation from 
\beq
\label{potential}
V(x,t)= q \left[U_1(t)\phi_1(x)+U_2(t)\phi_2(x)\right],
\eeq
where $U_i$ are segment voltages, $\phi_i$ are dimensionless electrostatic potentials, $q$ is the electric charge of the ion, and subindices $i=1,2$ correspond to the different segments. We use a ${}^{40}$Ca${}^+$ ion in the numerical calculations so $q$ is the elementary charge.

 
Electrostatic potentials are usually computed through the boundary element method or finite element method  solvers such as NIST BEM or COMSOL \cite{Singer2010,deClercq2015}, but the results can be well approximated by Gaussian 
functions  
$\phi_i=a \,e^{-(x-b_i)^2/2c^2}$,
see figure \ref{gaussiancurves}(c). The approximation provides analytical results and more exact but numerical
functions will not change quantitatively any of the conclusions drawn here.
%

%
%
%
%
%

By imposing that first and second derivatives of the potential $(\ref{potential})$  at $\alpha$
should be equal to $-F(t)$ and $m\omega^2$, respectively, we find 
\beq
U_i= \frac{(-1)^i m \omega^2 \phi_j'[\alpha(t)] +(-1)^i m \ddot{\alpha}(t) \phi_j''[\alpha(t)]}{\left\{ \phi''_2[\alpha(t)]\phi'_1[\alpha(t)] - \phi'_2[\alpha(t)] \phi''_1[\alpha(t)]\right\}q}, \hspace{0,8cm} i,j \in \{1,2\}, \hspace{0,2cm} j\neq i,
\eeq
where the primes  represent spatial derivatives. 
The same result may also be found as in \cite{Furst2014}, by splitting $U_i$ into two parts set to impose a harmonic potential term centered at $\alpha$ and a linear compensating term. 
%
%
%
%
%
%
%
 
%
%
%
\section{Results\label{results}}
%
%
%
%
%
\subsection{Energy and instantaneous power of the PS} 
The time-dependent energy of the ground dynamical mode $\psi_0(x,t)$ driven by $H$ is
\beq
\label{energyps}
E_{PS}=\frac{\hbar \omega}{2}+\frac m 2 \dot{\alpha}(t)^2 - m\ddot{\alpha}(t) \alpha(t) + f(t).
\eeq
%
Expanding the potential (\ref{potential}) in Taylor series
around $\alpha(t)$,  the additional time-dependent term in (\ref{compensatingforcepotential}) is given by  
%
\beq
\label{additionalterm}
f(t)=m\alpha(t)\ddot{\alpha}(t) - \frac{c^2m[c^2 \omega^2+(d-2\alpha(t))\ddot\alpha(t)]}{c^2-d\alpha(t)+\alpha(t)^2}, 
\eeq
which depends on the chosen reference trajectory $\alpha(t)$ and its acceleration, on the width of the Gaussian electrostatic potentials $c$, and on the 
distance $d$ between segment centers. 
Thus the power for the primary system, and any definition of energy consumption that depends on the energy of the PS, in fact 
depend on the control system through $f(t)$. Obviating the control system and thus leaving $f(t)$ indeterminate makes the energy of the PS undefined. 
Often $f(t)$  is taken as zero for simplicity,  
but this provides the wrong energy
function $E_{PS}(t;f=0)$, since it is not defined with respect to a fixed zero of energy so it cannot provide the true power.   
%

%
\begin{figure}[h]
\begin{center}
\includegraphics[width=0.45\textwidth]{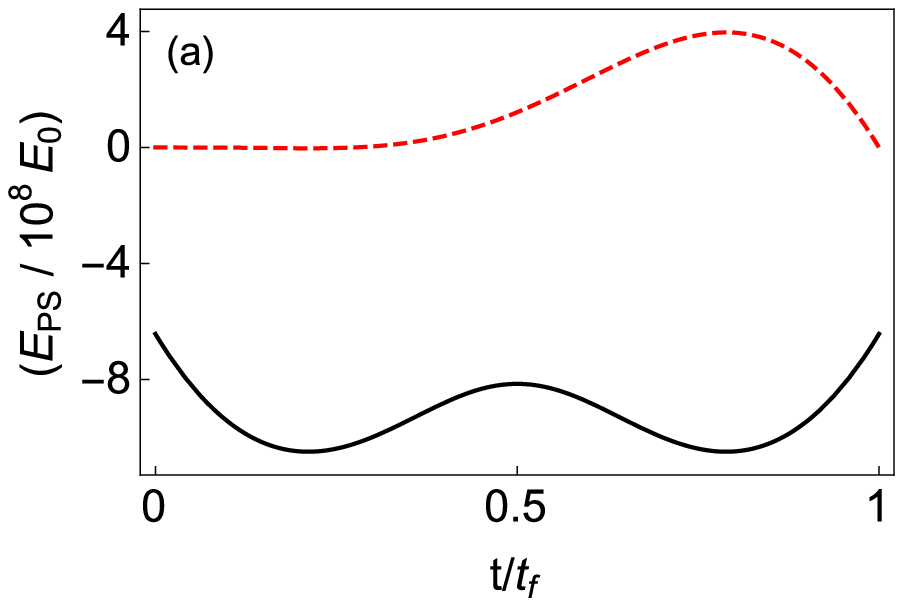} \,\,
\includegraphics[width=0.45\textwidth]{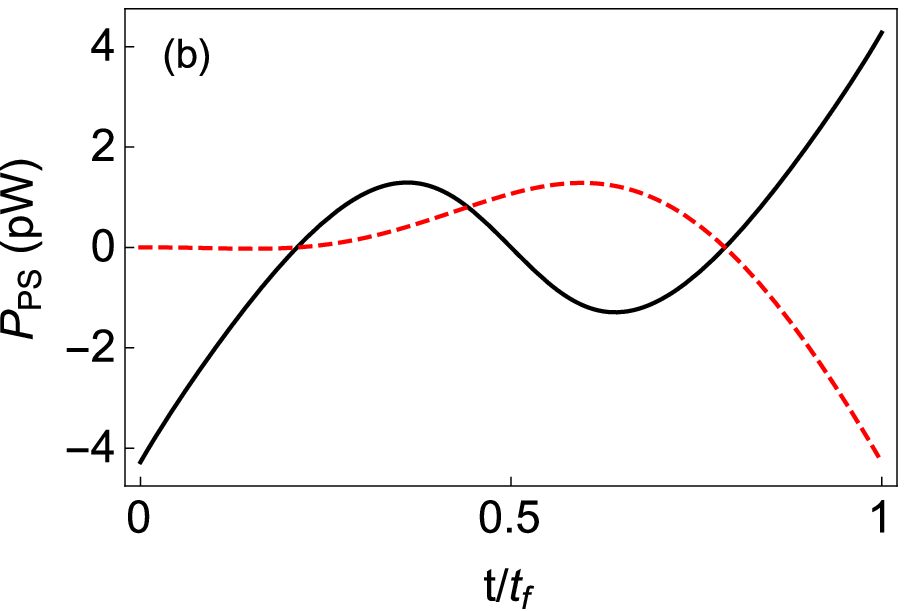}
\caption{(a) Scaled ratio between $E_{PS}$ in (\ref{energyps}) and the ground energy of the reference harmonic potential, $E_0=\hbar \omega / 2$. We consider a transport process of a ${}^{40}$Ca${}^+$ ion carried out by the potential (\ref{compensatingforcepotential}) with $f$ given by (\ref{additionalterm}) (solid black line), and  with $f=0$ (dashed red line), for  $\alpha(t)$ in (\ref{alfa}). (b) Corresponding power computed as the rate of $E_{PS}$ change. $t_f=0.418$ $\mu$s, $d = 280$ $\mu$m, and $\omega=2\pi\times1.3$ MHz.  \label{evolution_PS}}
\end{center}
\end{figure}
%
Figure \ref{evolution_PS}(a) depicts the completely different time evolution of the energy of the ion with the physical $f(t)$ in (\ref{additionalterm}) and with  $f=0$, and figure \ref{evolution_PS}(b) their corresponding instantaneous power, which for the physical $f(t)$ reads
\beq
\label{powerps}
P_{PS}=\frac{dE_{PS}}{dt} = m(A+B\omega^2),
\eeq
where $A$ and $B$  depend on the parameters of the trap
$c$ and $d$, and on the transport function $\alpha$ and its derivatives,  
\beqa  
\label{parameterspowerps}
\hspace*{-2cm}A=\dot{\alpha}(t)\ddot{\alpha}(t) - \frac {c^2 [d - 2\alpha(t)]^2 \dot{\alpha}(t)\ddot{\alpha}(t)}{[c^2 - d \alpha(t) + \alpha(t)^2]^2}
- \frac {c^2\{-2 \dot{\alpha}(t)\ddot{\alpha}(t)
 + [d-2\alpha(t)] \alpha^{(3)}(t)\}}{c^2 - d \alpha(t) + \alpha(t)^2}, 
\nonumber\\
\hspace*{-2cm}B=\frac{c^4 [-d\dot{\alpha}(t) + 2 \alpha(t) \dot{\alpha}(t)]}{[c^2 - d \alpha(t) + \alpha(t)^2]^2}.
\eeqa
Although not appreciable in the scale of the figure, at boundary times ($t_b = 0, t_f$)
$E_{PS}(t_b;f=0)$ 
is the ground state energy of the harmonic potential $E_0=\hbar\omega/2$,
while $E_{PS}(t_b)=E_0 - m \omega^2 c^2$, that is, initial and final energies have been displaced from the ground energy of the reference harmonic potential by $f(t_b)$. Both processes, with $f=0$ or $f(t)$ given by (\ref{additionalterm}), are formally valid shortcuts without final excitations on the transported state, and, seemingly, with no energy cost as the power $P_{PS}$ integrates to zero in both cases. This is a general property in STA processes with the same energy of the PS at boundary times, as in transport protocols. The instantaneous power does not integrate to zero in STA processes that imply a net energy change for the PS, such as expansions or compressions. 

%
%
%
%
%
%
%
%
%
%
%

\subsection{Power of the CS} 

\begin{figure}
\begin{center}
\includegraphics[width=0.41\textwidth]{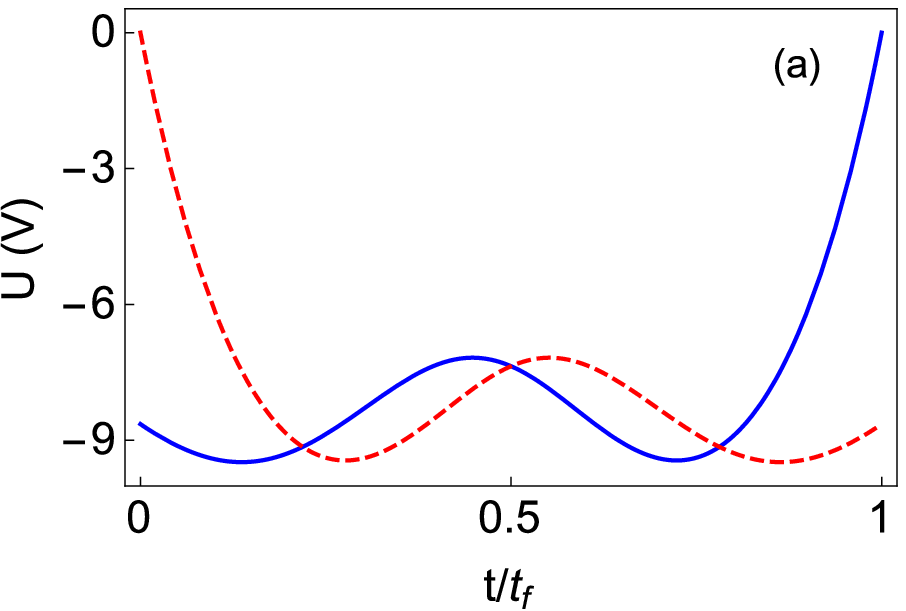} 
\includegraphics[width=0.425\textwidth]{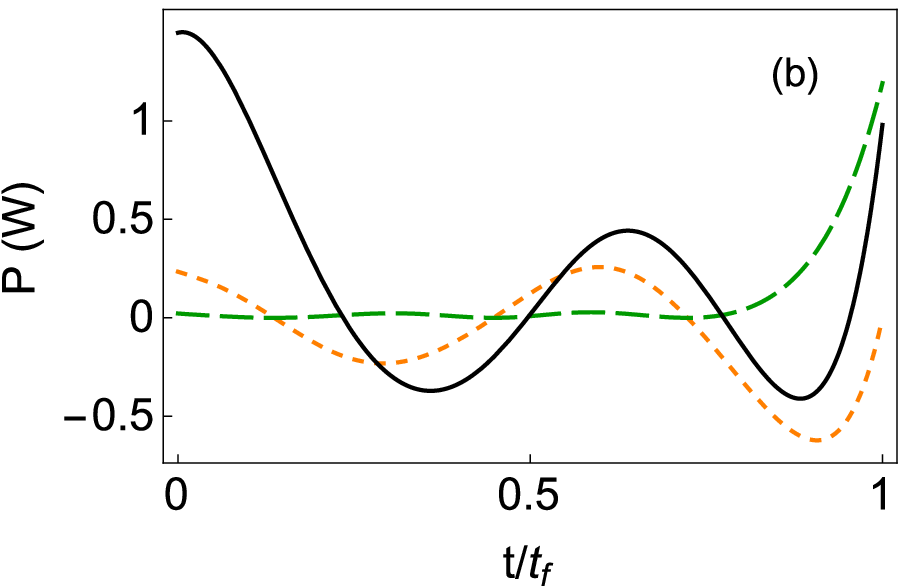}
\caption{(a) Time evolution of the voltage $U_1$ (blue solid line) of segment one, and the voltage $U_2$ (red dashed line) of segment two. 
Note the reflection symmetry. They generate the potential (\ref{potential}) with  
$\omega=2\pi\times1.3$ MHz that controls the transport of a ${}^{40}$Ca${}^+$ ion in a segmented Paul trap over a distance $d=280$ $\mu$m in $t_f=0.418$ $\mu$s. (b) Power to charge the capacitor, $P_{C_1}$ (short-dashed orange line), with $C=1$ nF, and power dissipated in the resistance, $P_{R_1}$ (long-dashed green line), with $R=30$ $\Omega$, see (\ref{powerscontrol}). The corresponding powers for segment two are reflection-symmetric with respect to the middle time. The total power 
$P_{CS}$ is also shown (solid black line).  \label{voltageandpowertermsCS}}
\end{center}
\end{figure}

Let us now consider 
the power we have to supply to the 
CS to implement the STA protocol. Transporting an ion requires moving the potential minimum by varying the segment voltages  in time, as explained in Methods. 
This is achieved by inducing currents that go through the RC low-pass filters and
govern the voltages in the electrodes. The total power exerted by the electromotive force at the source of the electrode circuits includes the rate of change of the energy accumulated at the capacitor and the power dissipated in the resistance through the Joule effect, 
$P_{CS}=\sum_i (P_{C_i}+P_{R_i})$, 
respectively given by 
\beq
\label{powerscontrol}
P_{C_i}=C \, U_i \partial_t U_i, \hspace{2cm}
P_{R_i} = R \, C^2 \, (\partial_t U_i)^2,
\eeq
%
where \textit{R} and \textit{C} are the resistance and the capacitance of each electrode circuit (assumed equal for both segments). See figure \ref{voltageandpowertermsCS}(b) for the evolution of these power terms in the first segment.  
Those for the second segment are symmetrical. 

The power required by the CS is orders of 
magnitude larger than the one for the PS, as we are dealing with macroscopic charges instead of a single ion. In fact this
disparity of scales is helpful in that the effect of the exact state of the PS has a negligible influence in the implementation of the protocol. This is one of the observations in  the mechanical crane model in \cite{Torrontegui2017}, where  the stability of the STA protocol in the control system required a small mass of the load compared to the mass of the trolley. (Otherwise each initial condition of the PS would require a different control protocol.) 

\subsection{Comparison between the energy consumed by the PS and by the CS}
In \cite{Torrontegui2017} it was emphasized that the way to implement a negative power has a decisive influence on the energy cost. Negative powers do not necessarily imply 
a reduction in the energy cost of the process. To implement such a reduction, the system has to store and reuse the energy given away, which is often not the case or only partially true. 
To calculate the energy consumption by integration of the power, Torrontegui et al. proposed to include a parameter  $\eta$ in the negative power segments, 
%
\beq
\label{energywitheta}
\pazocal{E}=\int_0^{t_f} P_+ dt + \eta \int_0^{t_f} P_- dt.
\eeq
Here $P_{\pm} = \Theta(\pm P) P$ are positive/negative parts of the power of the system and $-1\le \eta \le1$ accounts for different possible 
scenarios. The limit $\eta=-1$ means that the negative power implies as much energy consumption as the positive one, while $\eta=1$
means that the energy can be stored and reused (regenerative braking).  

%
%
\begin{figure}[h]
\begin{center}
\includegraphics[width=0.415\linewidth]{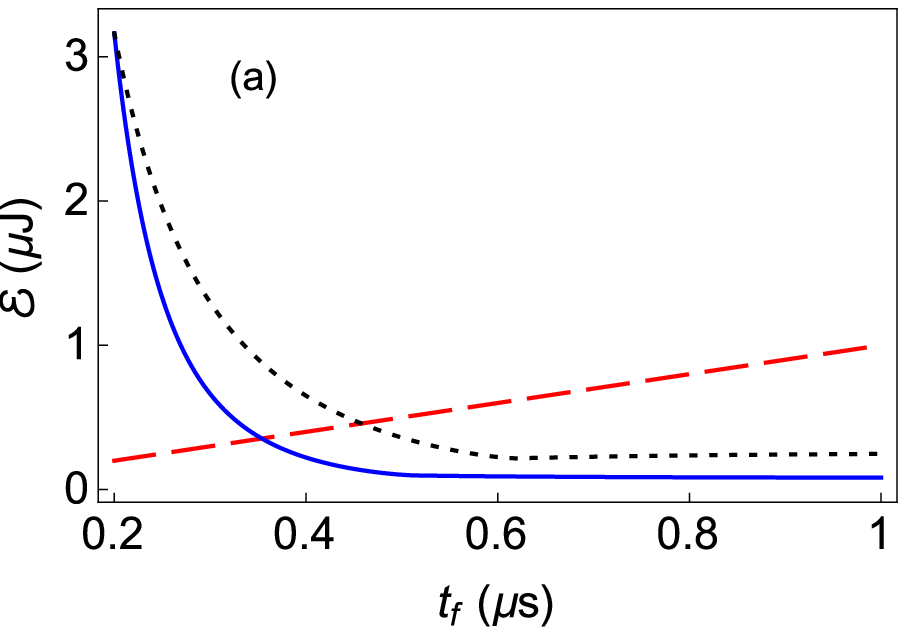}
\includegraphics[width=0.43\textwidth]{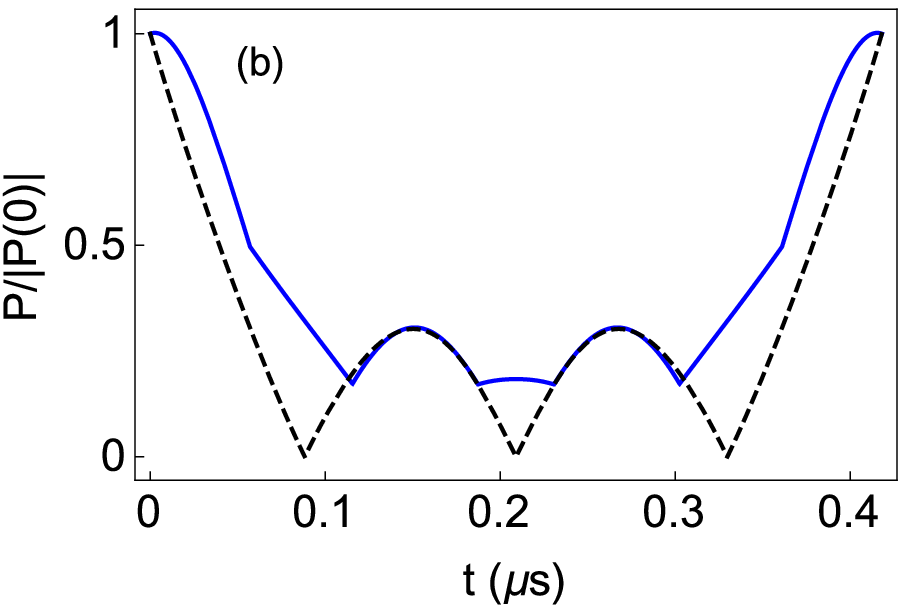}
\caption{ (a)  Energy consumption in the electrode circuits ($\pazocal{E}_{CS}$) that control the transport of a ${}^{40}$Ca${}^+$ ion  accelerated by a compensating force STA for different final times  (solid blue line);  scaled energy consumed by the ion ($\gamma \pazocal{E}_{PS}$) in such process, where the factor $\gamma=10^{11}$ is set so that the starting point of both curves coincide (short-dashed black line);  energy cost of inducing the rf-fields that confine the ion assuming that it requires a constant power of $P=1$ W  (long-dashed red line). Parameters:  $d=280$ $\mu$m,  $\omega=2\pi\times1.3$ MHz,
$R=30$ $\Omega$, and $C=1$ nF.  (b) Absolute value of the power consumed by the ion $|P_{PS}|$ (dashed black line) and power required by the control computed as $\sum_{i=1,2} |P_{C_i}| + P_{R_i} $ (solid blue line), for  the same process with $t_f=0.418$ $\mu$s. 
Curves have been scaled for better comparison, dividing them by their absolute value at initial time ($|P_{CS}(0)| = 1.45$ W and $|P_{PS}| = 4.28\times 10^{-12}$ W). \label{evolution_power_csps}}
\end{center}
\end{figure}
%
%

In our CS, the power dissipated in the resistance, $P_{R_i}$, is always positive but $P_{C_i}$ becomes negative when the capacitor is discharging, see figure \ref{voltageandpowertermsCS}(b). However, the short time scales intrinsic to STA require to charge and discharge the capacitor much faster than the circuit's time constant, so we always need to actively drive it. This is achieved by changing the polarity of the power source, reversing the direction of the current whenever we need to change the energy flow in the capacitor. This makes impossible to retrieve the energy stored in the capacitor, which translates as an $\eta = -1$ scenario in our analysis.
The energy consumed by the control is in summary given by
\beq
\label{realenergyCS}
\pazocal{E}_{CS}= \int_0^{t_f} \sum_{i=1}^2 \,(|P_{C_i}| + P_{R_i}) \,dt.
\eeq
For the reference trajectory (\ref{alfa}) and $t_f\to 0$
it scales as $t_f^{-5}$ ($\sim t_f^{-4}$ for the time scale considered  in figure \ref{evolution_power_csps}(a)) while $\pazocal{E}_{PS}$, 
defined as 
\beq
\label{realenergyPS}
\pazocal{E}_{PS}=\int_0^{t_f} | P_{PS}| dt
\eeq
by analogy with  $\pazocal{E}_{CS}$,
scales as $t_f^{-2}$ (see figure \ref{evolution_power_csps}(a)). 

In figure \ref{evolution_power_csps}(b) we compare $|P_{PS}|$ and $\sum_{i=1,2} |P_{C_i}| + P_{R_i}$ normalized 
by their initial value.  
Although they have a similar evolution, they are not just scaled with respect to each other. 
A consequence of their different orders of magnitude is that the actual energy consumption
$\pazocal{E}$ can be computed in terms of the CS alone with great accuracy, i.e.,
$\pazocal{E}=\pazocal{E}_{PS} + \pazocal{E}_{CS} \approx \pazocal{E}_{CS}$.
\subsection{Power peaks}
\begin{figure}[h]
\begin{center}
\includegraphics[width=0.5\linewidth]{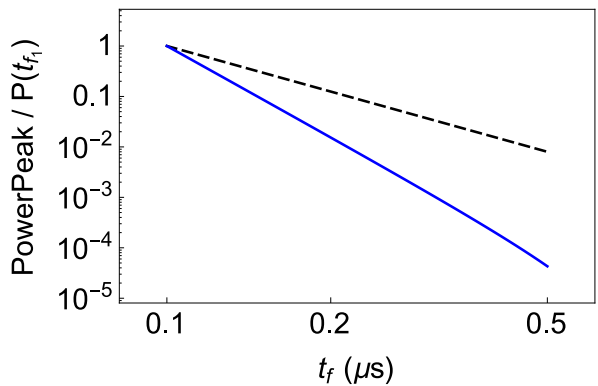}
\caption{Scaled power peaks required for shuttling a ${}^{40}$Ca${}^+$ ion over $d=280$ $\mu$m for different final times. Dashed black line corresponds to the PS and solid blue line to the CS. They are normalized to one at $t_{f1}=0.1$ $\mu$s. The specific values are $P_{PS}(t_{f_1}) = 3.13 \times 10^{-10}$ W and $P_{CS}(t_{f_1}) = 6455$ W. Other parameters used are $R=30$ $\Omega$, $C=1$ nF and $\omega=2\pi \times 1.3$ MHz. }
\label{powerpeaks}
\end{center}
\end{figure}

The power peaks of the protocol may limit the minimum time to implement a STA, 
as a generic power source is only able to reach a certain 
maximum value. Figure \ref{powerpeaks} depicts the value of the power peak of the PS and the CS for different final times in a transport process with 
(\ref{alfa}).    
For the CS the power peak occurs at  boundary times ($t_b = 0, t_f$) and it reads
\beq
P_{CS}(t_b) =  \frac{m^2}{q^2}\left(\frac{G}{t_f^6} + \frac{ J \omega^2}{t_f^{3}}\right),
\eeq
where 
\beq
G=\frac{3600 R C^2}{a^2} [(c^2-d^2)^2 + c^4 e^{(d/c)^2} ]; \hspace{1cm} J=\frac{60Cc^2}{a^2}(c^2 - d^2).
\eeq
For the PS and for the parameters used in the paper and final times shorter than $1$ $\mu$s, the power peak is at initial and final times as well (see figure \ref{evolution_power_csps} where $t_f=0.418$ $\mu$s) and it is given by 
\beq
P_{PS}(t_b) = \frac{60 d^2 m}{t_f^3}.
\eeq
Again, the difference between the PS and the CS power peaks is not just a matter of scaling, they show a different  qualitative behavior. The power peak of the PS scales as $t_f^{-3}$ while the one of the CS scales as $t_f^{-6}$. 
%
\subsection{Optimization\label{optim}}
The freedom to choose different transport functions $\alpha(t)$ may be used to optimize physically relevant variables. For example this freedom was used in \cite{Gonzalez-Resines2017}  to avoid deviations from the harmonic regime at intermediate times in the transport of a load by a mechanical crane.
Here we use a $7^{th}$ degree polynomial ansatz with two free parameters to minimize the energy consumption of the transport process. 
%
The optimization, i.e., the final form of the reference trajectory, must be based on minimizing the total energy consumption. 
However, it is interesting to compare the results with alternative optimization criteria. In particular, 
we shall also minimize the energy consumption of the primary system $\pazocal{E}_{PS}$ with the physical, CS-based $f$ function  (\ref{additionalterm}), and $\pazocal{E}_{PS(f=0)}$ with $f=0$.  
Table  \ref{tab_energy_opt}  shows  the energy consumption of the PS
(with $f=0$ and with the physical $f(t)$), and of the CS  for each of the optimized protocols. Notice that the optimization of $\pazocal{E}_{PS}$  with the physical $f$, yields essentially the same  results than the optimization of $\pazocal{E}_{CS}$. Both protocols achieve a reduction in the consumption of $\%6$ for the ion and $\%17$ for the control. On the contrary, optimizing $\pazocal{E}_{PS(f=0)}$ 
turns out to be unsatisfactory, as it increases significantly the total energy consumption of the process: $\pazocal{E}_{PS}$ increases by $\%13$ and $\pazocal{E}_{CS}$ by $\%43$ with respect to the non-optimized trajectory that uses the fifth degree polynomial (\ref{alfa}).  These numerical results 
confirm the importance of including the $CS$-dependent term $f$ in the $PS$-energy so as to mimic the evolution of the energy consumption and to use $PS$-energies for optimizing consumptions.      
  

\begin{table}[t]
\centering
\begin{tabular}{|c|c|c|c|c|}

\hline
 & & & & \\
 & \textbf{Non optimized} & \textbf{Optimized for } & \textbf{Optimized for}  & \textbf{Optimized for} \\
 & & $\pazocal{E}_{PS(f=0)}$ & $\pazocal{E}_{PS}$ & $\pazocal{E}_{CS}$  \\
 
\hline

$\pazocal{E}_{PS(f=0)} $  & $ 3.441 \times 10^{-19}$ J & $ 2.179 \times 10^{-19}$ J & $ 3.363 \times 10^{-19}$ J & $ 3.355 \times 10^{-19}$ J\\

\hline

$\pazocal{E}_{PS} $ & $ 5.513 \times 10^{-19}$ J & $ 6.242 \times 10^{-19}$ J & $ 5.169 \times 10^{-19}$ J & $ 5.169 \times 10^{-19}$ J\\

\hline

$\pazocal{E}_{CS}$ & $1.882 \times 10^{-7}$ J & $ 2.696 \times 10^{-7}$ J & $ 1.572 \times 10^{-7}$ J & $ 1.572 \times 10^{-7}$ J\\

\hline
\end{tabular}
\caption{\label{tab_energy_opt} Energy consumptions for transport processes with reference trajectory given by $\alpha(t) = \sum_{j=0}^7 a_j (t/t_f)^j$. $\pazocal{E}_{PS(f=0)}$ is the energy consumed by the ion given in (\ref{realenergyPS})  with $f(t)=0$, $\pazocal{E}_{PS}$ is the same quantity  with $f(t)$ given by (\ref{additionalterm}) and $\pazocal{E}_{CS}$ is the energy consumption in the control given in (\ref{realenergyCS}).  Parameters $a_j$, $j<6$, are fixed (as functions of $a_6$ and $a_7$) by the boundary conditions described in Methods. The first titled column corresponds to the original non-optimized protocol with (\ref{alfa}),
while the others correspond to different criterions to find the free parameters, based on minimizing one of the mentioned energy consumptions. The values of the free parameters are $a_6 = -0.0195$ and $a_7 = -0.0049$ for the "Optimized for $\pazocal{E}_{PS(f=0)}$" column, $a_6 = -0.0093$ and $a_7 = 0.0027$ for the "Optimized for $\pazocal{E}_{PS}$" column and $a_6 = -0.0094$ and $a_7 = 0.0027$ for the "Optimized for $\pazocal{E}_{CS}$" column. } 
\end{table}

\section{Discussion}
As new quantum technologies unfold from laboratory prototypes to commercially available devices, energetic costs 
of processes may become more and more relevant. Shortcuts to adiabaticity can play an important role in this 
transition by providing a toolbox of approaches to design control protocols that minimize process times and
the effects of decoherence. 
Determining the energetic cost of a shortcut requires a global perspective that includes the primary system and the 
control system as well. The shortcuts are by definition fast processes so one cannot assume that the control system may  
change infinitely slowly to avoid dissipation, as in Landauer's analysis of minimal costs of 
computation \cite{Landauer1961}, 
 or in ideal thermodynamical reversible processes. To be more precise, very slow processes are physically possible, but 
STA are never applied in the long-time domain.

The study case chosen in this paper is a microscopic ion transported with a STA protocol implemented
by macroscopic operations to modulate the voltages of a segmented Paul trap. Features of the energy consumption that were speculated to be broadly applicable 
after the analysis of a mechanical crane \cite{Torrontegui2017} have been found here too.  For example, negative power time-segments may imply 
as much consumption as the positive power segments.  In the model, as it will be typically the case in controlling microscopic systems, 
the consumption is dominated by far by the control system. This is in fact desirable,  
otherwise the control operations to implement a given STA  would have to depend on the specific initial conditions of the PS.  
The power for the CS is due to dissipation in the resistances and to the charge or discharge of capacitors. This dual origin
(dissipative and non-dissipative)  
is once again analogous to the 
mechanical crane model, where the power was employed to compensate dissipation (friction losses) and move (accelerate or brake) the control trolley.      

The integrated energy consumption of the PS alone is not zero when evaluated with the absolute value of the power of the PS. 
This integral quantity,  properly scaled,  resembles the consumption of the CS, and in fact 
can be used  to find optimal transport trajectories, but only when a purely time-dependent energy shift that depends on the CS 
is included in the PS Hamiltonian. In other applications this term is neglected or set as zero, 
but it is a crucial factor  to determine energy flows.   

We have paid attention to global energy consumptions rather than to differential ones (relative to some reference process). 
A definition of energy consumption based on a differential power may have some uses, e.g., to compare different ways to 
achieve a shortcut for a given  reference process. However, it depends on the reference process
and it is inappropriate if we are interested in the actual 
energy consumption, the reason being that the reference process also consumes energy.  
     
%
The main text has focused on the nontrivial part of the energy consumption of the CS, associated with the dc electrodes, which grows strongly 
when diminishing process times,  leaving aside 
the linear-in-time consumption of the rf-electrodes.  
Combining the two contributions, minimal times for energy consumption can be identified. 

Further examples of systems subjected to STA may be examined to build a general theory, e.g. analyzing energy consumptions in
discrete systems \cite{Hu2018}.

\ack
We thank Kihwan Kim for discussions. 
We acknowledge    
funding by 
the Basque Government (Grant No. IT986-16) 
and MINECO/FEDER,UE (Grant No. FIS2015-67161-P).
The research is based upon work supported by the Office of the Director of National Intelligence (ODNI), Intelligence Advanced Research Projects Activity (IARPA), via the U.S. Army Research Office grant W911NF-16-1-0070. The views and conclusions contained herein are those of the authors and should not be interpreted as necessarily representing the official policies or endorsements, either expressed or implied, of the ODNI, IARPA, or the U.S. Government. The U.S. Government is authorized to reproduce and distribute reprints for Governmental purposes notwithstanding any copyright annotation thereon. Any opinions, findings, and conclusions or recommendations expressed in this material are those of the author(s) and do not necessarily reflect the view of the U.S. Army Research Office.
\vspace*{1cm}

\bibliography{Bibliography}{}
\bibliographystyle{sofia}

\end{document}